\documentclass[aps,prc,superscriptaddress,showpacs,floatfix,nofootinbib,notitlepage,twocolumn]{revtex4-1}
\usepackage{amsmath,graphicx,float,hyperref,upgreek}
\newcommand{\trento}{T$\mathrel{\protect\raisebox{-2.1pt}{R}}$ENTo}
\def\bra{\langle}
\def\ket{\rangle}

\begin{document}

\title{Skewness of mean transverse momentum fluctuations in heavy-ion collisions}

\author{Giuliano Giacalone}
\affiliation{Universit\'e Paris Saclay, CNRS, CEA, Institut de physique th\'eorique, 91191 Gif-sur-Yvette, France}
\author{Fernando G. Gardim}
\affiliation{Instituto de Ci\^encia e Tecnologia, Universidade Federal de Alfenas, 37715-400 Po\c cos de Caldas, MG, Brazil}
\author{Jacquelyn Noronha-Hostler}
\affiliation{Department of Physics, University of Illinois at Urbana-Champaign, Urbana, IL 61801, USA}
\author{Jean-Yves Ollitrault}
\affiliation{Universit\'e Paris Saclay, CNRS, CEA, Institut de physique th\'eorique, 91191 Gif-sur-Yvette, France}

\begin{abstract}
We propose the skewness of mean transverse momentum, $\bra p_t \ket$, fluctuations as a fine probe of hydrodynamic behavior in relativistic nuclear collisions. We describe how the skewness of the $\bra p_t \ket$ distribution can be analyzed experimentally, and we use hydrodynamic simulations to predict its value. We predict in particular that $\bra p_t \ket$ fluctuations have positive skew, which is significantly larger than if particles were emitted independently. We elucidate the origin of this result by deriving generic formulas relating the fluctuations of $\bra p_t \ket$ to the fluctuations of the early-time thermodynamic quantities. We postulate that the large positive skewness of $\bra p_t \ket$ fluctuations is a generic prediction of hydrodynamic models.
\end{abstract}

\maketitle

\section{Introduction} 
In ultrarelativistic nucleus-nucleus collisions, the mean transverse momentum, $\langle p_t\rangle$, of emitted particles fluctuates event to event, for a given collision centrality. 
There are trivial \textit{statistical} fluctuations of $\bra p_t \ket$, due to the fact that the average is evaluated over a finite sample of particles, but the observed fluctuations are larger. 
The excess fluctuations are called \textit{dynamical} fluctuations, and have been measured in Au+Au collisions at $\sqrt{s_{\rm NN}}=200$~GeV~\cite{Adams:2003uw} and lower energies~\cite{Adams:2005ka,Adamczyk:2013up,Adam:2019rsf}, and in Pb+Pb collisions at $\sqrt{s_{\rm NN}}=2.76$~TeV~\cite{Abelev:2014ckr}. 
In hydrodynamic models of particle production, dynamical $\langle p_t\rangle$ fluctuations originate from event-to-event fluctuations at the early stage of the collision~\cite{Broniowski:2009fm,Bozek:2012fw}. 
$\langle p_t\rangle$ fluctuations have received much less attention in hydrodynamic studies than anisotropic flow~\cite{Gale:2012rq,Heinz:2013th}, yet they are a more direct way of observing initial-state fluctuations. 
They actually strongly constrain the modeling of the initial stages, and only a few recent hydrodynamic studies are able to reproduce experimental data on $\langle p_t\rangle$ fluctuations~\cite{Bozek:2017elk,Bernhard:2019bmu,Everett:2020xug}.

In this paper, we argue that, at a given collision centrality, the probability distribution of $\langle p_t\rangle$ is not Gaussian, but has positive skew. 
In Sec.~\ref{s:evidence} we show that a hint of this positive skew can be seen in existing STAR data~\cite{Adams:2005ka} on Au+Au collisions, while it is clearly visible in the results of event-by-event hydrodynamic simulations of Pb+Pb collisions.
This motivates us to investigate this phenomenon. 
We define measures of the skewness of $\langle p_t\rangle$ fluctuations in Sec.~\ref{s:measuring}, with detailed explanations about the analysis procedure to measure them given in Appendix~\ref{s:cumulants}, and we make quantitative predictions for these quantities using hydrodynamic calculations in Sec.~\ref{s:hydro}.
The resulting skewness is significantly larger than if particles were independent. 

We investigate, hence, the origin of the skewness.
In Sec.~\ref{s:initial}, we use the idea put forward in Refs.~\cite{Gardim:2020sma,Giacalone:2020dln} that the fluctuations of $\langle p_t\rangle$ at a given centrality originate from the fluctuations of the total energy in the fluid at the initial condition, $E_0$. We first show that the distribution of $E_0$ is indeed positively skewed in our hydrodynamic calculation, and then argue that this is likely to be observed in any hydrodynamic calculation. 
This is done in Sec.~\ref{s:perturbative}, where we derive a generic formula relating the skewness of the $E_0$ distribution to the statistical properties of the initial density field in a perturbative approach~\cite{Blaizot:2014nia,Floerchinger:2014fta}.

\begin{figure}[b]
    \centering
    \includegraphics[width=.85\linewidth]{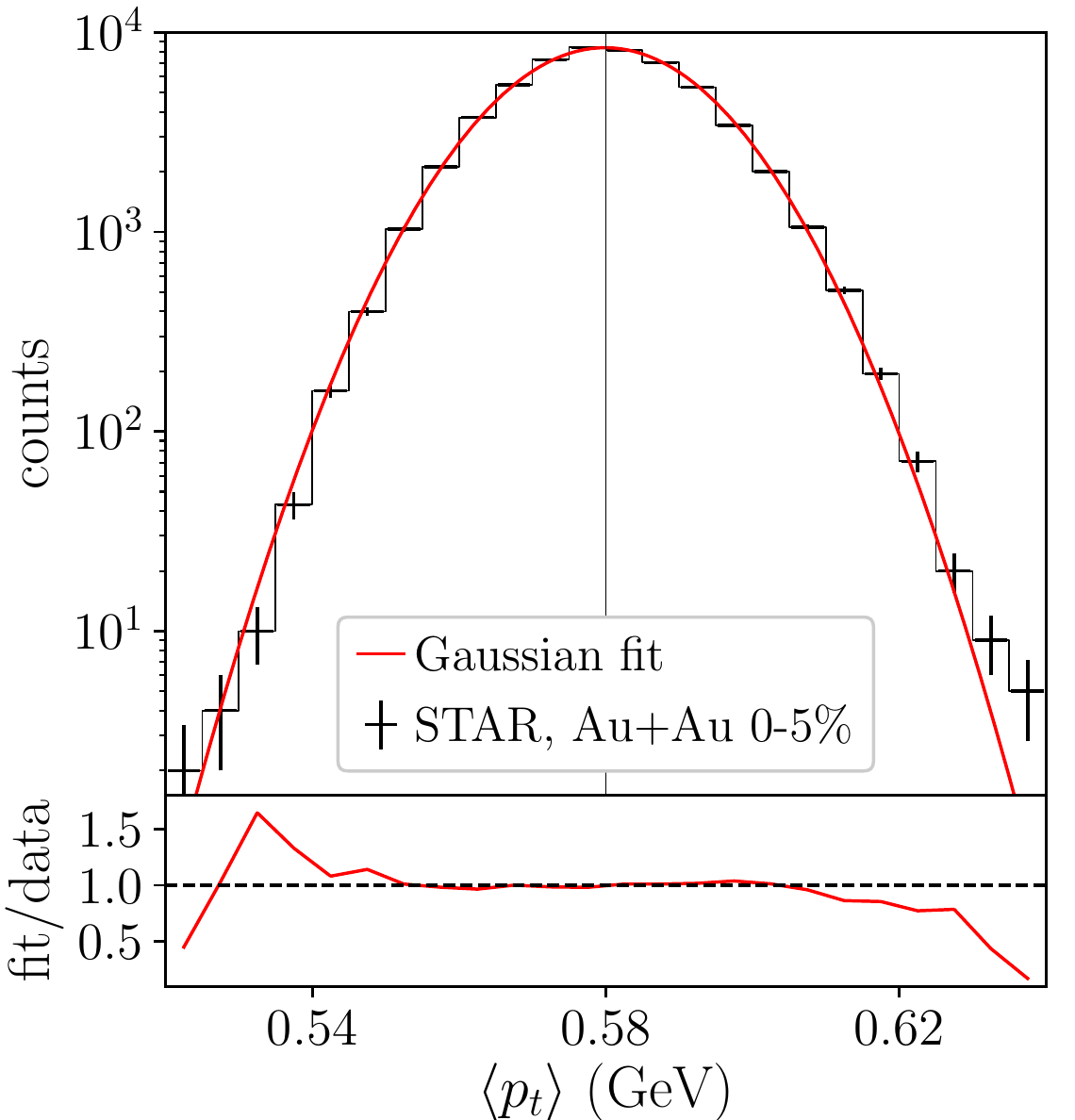}
    \caption{(Color online) Distribution of $\langle p_t\rangle$ for Au+Au collisions at $\sqrt{s_{\rm NN}}=200$~GeV in the 0-5\% centrality window. Data from the STAR collaboration~\cite{Adams:2005ka} are shown as a histogram.  The solid line is a Gaussian fit to these data. The lower panel is the ratio between the Gaussian fit and the data. The data are above the Gaussian to the right, and below the Gaussian to the left, which hints at a positive skew.}
    \label{fig:star}
\end{figure}

\section{Skewness in data and in hydrodynamics}
\label{s:evidence}

Figure~\ref{fig:star} displays the histogram of the distribution of  $\langle p_t\rangle$ measured by the STAR collaboration in central Au+Au collisions~\cite{Adams:2005ka}, where $\langle p_t\rangle$ is evaluated by averaging the transverse momenta of the charged particles observed in the detector. 
As mentioned in the Introduction, this quantity has trivial fluctuations due to the finite number of particles, typically of order 1000, in every event. 
The width of the distribution of $\langle p_t\rangle$ is actually dominated by these statistical fluctuations, and the dynamical fluctuations only represent a modest fraction of this width. 
Even though this histogram does not represent a distribution of dynamical fluctuations, it is instructive to see that the distribution is not symmetric. Comparison with a Gaussian fit, shown as a solid line, shows that the data points are above the fit to the right, and below the fit to the left, which is an indication that the distribution of $\langle p_t\rangle$ has positive skew. 
A quantitative calculation gives for the standardized skewness $(7.3\pm 1.0)\times 10^{-2}$, which shows that the skew is far beyond error bars, even though the statistics is modest ($\sim 5\times 10^4$ events). 
We use this as an illustration that the skewness should be easy to measure accurately with the large statistics now available at colliders. 
However, the qualitative prediction that the skewness is positive is to some extent trivial when fluctuations are large. 
The reason is that the transverse momentum is positive by construction, which naturally produces the left-right asymmetry seen in Fig.~\ref{fig:star}. 
In particular, the same study applied to mixed events~\cite{Adams:2005ka}, made up artificially using particles from different events, results in a skewness of comparable magnitude, even though it contains no dynamical information by construction. 
Therefore, it is essential to isolate dynamical fluctuations before measuring the skewness, as will be explained in Sec.~\ref{s:measuring}. 
 
We present now the distribution of $\langle p_t\rangle$ in event-by-event hydrodynamics. 
We do not run new hydrodynamic calculations, but use results from a prior high-statistics simulation, in which 50000~minimum bias Pb+Pb collisions at $\sqrt{s_{\rm NN}}=5.02$~TeV were generated~\cite{Alba:2017hhe,Giacalone:2017dud}. 
This hydrodynamic calculation was shown to successfully reproduce the observed magnitude and centrality dependence  of anisotropic flow  ($v_2$, $v_3$, $v_4$), and to slightly overestimate the mean transverse momentum of charged particles $\langle p_t\rangle$. 
Back then, we had not evaluated $\langle p_t\rangle$ fluctuations in this hydrodynamic calculation. 
It turns out that it overestimates their magnitude. 
This is a common limitation of event-by-event hydrodynamic calculations~\cite{Bozek:2012fw,Gardim:2019iah}, which has been overcome recently by using smoother initial conditions~\cite{Bozek:2017elk,Bernhard:2019bmu,Everett:2020xug}.
Note that agreement with $v_3$ data, for which initial fluctuations are essential, then requires to model the nucleon substructure~\cite{Moreland:2018gsh,Nijs:2020ors}. 
Since our hydrodynamic calculation does not quantitatively reproduce the magnitude of  $\langle p_t\rangle$ fluctuations, our predictions for the skewness are also not fully quantitative, as will be discussed below. 

The setup of our hydrodynamic calculation is the following. 
We start from a boost-invariant initial profile of entropy density given, event-to-event, by the \trento{} model of initial conditions~\cite{Moreland:2014oya}, which has been tuned following Ref.~\cite{Bernhard:2016tnd}.\footnote{We use $p=0$, corresponding to a geometric average of nuclear thickness functions. The thickness of a nucleus is a linear superimposition of participant nucleon thicknesses, which are taken as Gaussian profiles of width $w=0.51$~fm. The normalization of each nucleon thickness fluctuates following a gamma distribution of unit mean and standard deviation $1/\sqrt{k}$, where we use $k=1.6$.}
Events are sorted into centrality bins according to their total initial entropy (5\% bins are used). This is done to mimic the centrality selection performed in experiments.
We neglect the pre-equilibrium dynamics of the system~\cite{Vredevoogd:2008id,vanderSchee:2013pia,Kurkela:2018wud}, which is evolved hydrodynamically starting from proper time $\tau_0=0.6$~fm/c after the collision~\cite{Kolb:2000fha}  through the viscous hydrodynamic code {\footnotesize V-USPHYDRO}~\cite{Noronha-Hostler:2013gga,Noronha-Hostler:2014dqa,Noronha-Hostler:2015coa}, 
We implement a small specific shear viscosity, $\eta/s=0.047$~\cite{Alba:2017hhe}, and the 2+1 equation of state from lattice QCD~\cite{Borsanyi:2013bia}.
Fluid elements hadronize~\cite{Teaney:2003kp} when reaching a temperature of $150$~MeV. We include all hadronic resonances in the freezeout process (from the PDG16+ list \cite{Alba:2017mqu}), and their subsequent strong decays, but we neglect rescattering in the hadronic phase~\cite{Bass:2000ib,Teaney:2001av,Bernhard:2016tnd}.

Each hydrodynamic ``event'' corresponds to a different initial condition~\cite{Aguiar:2001ac,Holopainen:2010gz,Petersen:2010cw,Schenke:2010rr}.
The output of hydrodynamics is the continuous probability distribution of the transverse momentum~\cite{Cooper:1974mv,Mazeliauskas:2018irt}, which one integrates to calculate the mean value, $\langle p_t\rangle$. 
Therefore, the statistical fluctuations mentioned in the discussion of Fig.~\ref{fig:star}, due to the finite event multiplicity, are absent in the hydrodynamic calculation, so that the event-to-event fluctuations of $\langle p_t\rangle$ are the dynamical fluctuations themselves. 
The histogram of the distribution of $\langle p_t\rangle$ is displayed as solid lines in Fig.~\ref{fig:hydro} for two different centrality windows.     
Note that the values of $\langle p_t\rangle$ are larger than in Fig.~\ref{fig:star}, because the collision energy is much higher.\footnote{Also, our calculation overestimates $\langle p_t\rangle$ by a few percent even at the higher energy, as discussed in Ref.~\cite{Giacalone:2017dud}.} 
As mentioned above, our model of initial conditions overestimates $\langle p_t\rangle$ fluctuations, and the width in Fig.~\ref{fig:hydro} is too large by a factor $\sim 2$. 
Our point here is that the distributions of $\langle p_t\rangle$ in Fig.~\ref{fig:hydro} are clearly asymmetric, with a long tail on the right. This positive skew is more pronounced in peripheral collisions [panel (b)] than in central collisions [panel (a)]. 
However, the comment made about Fig.~\ref{fig:star} also applies here: 
Namely, the condition that  $\langle p_t\rangle$ is positive naturally generates a positive skewness, also for dynamical fluctuations. 
It is therefore essential to define a baseline corresponding to the value of the skewness naturally generated by the positivity condition. 
This is a non-trivial issue, which will be discussed in Sec.~\ref{s:baseline}.

\section{Measuring the skewness}
\label{s:measuring}

\begin{figure*}[t]
    \centering
    \includegraphics[width=\linewidth]{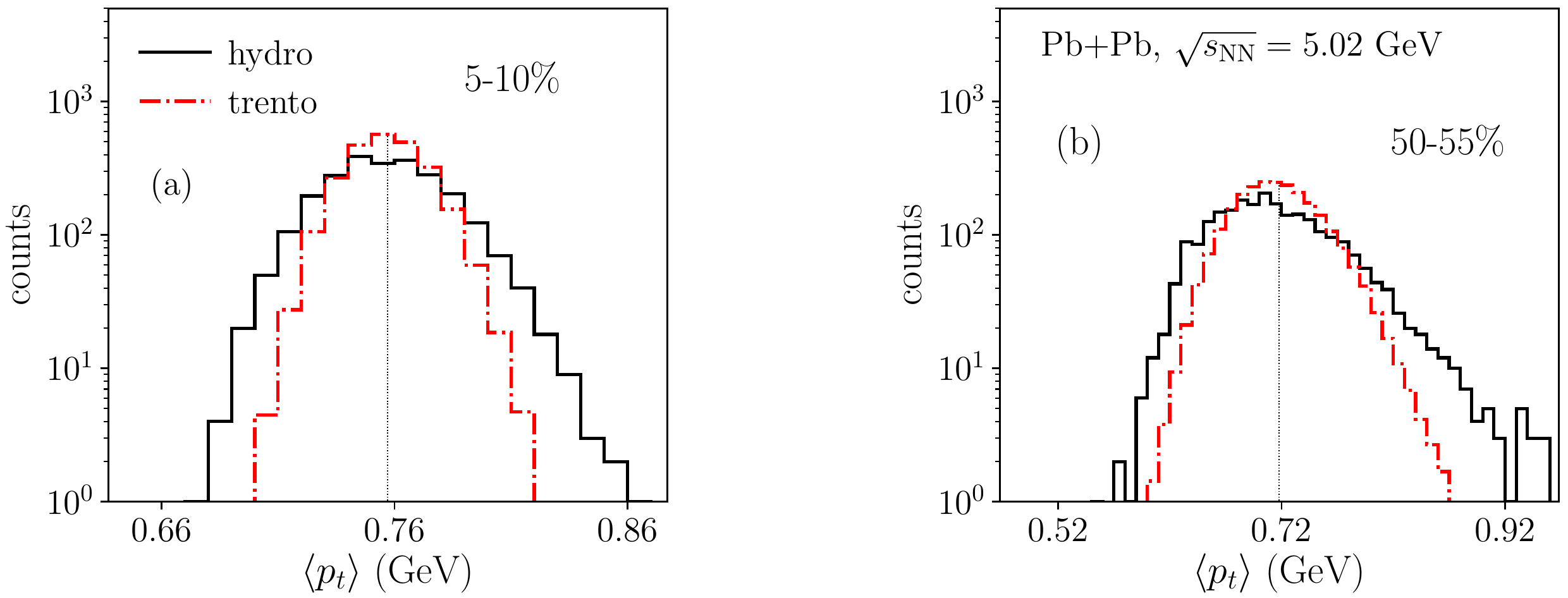}
    \caption{(Color online) Solid lines: Distribution of $\langle p_t\rangle$ in event-by-event hydrodynamic simulations of Pb+Pb collisions at $\sqrt{s_{\rm NN}}=5.02$~TeV~\cite{Giacalone:2017dud}, for charged particles in the transverse momentum interval $0.2<p_t<3$~GeV and in the pseudorapidity interval $|\eta| <0.8$. 
    Dash-dotted line: Distribution of $E_0/S$, where $E_0$ and $S$ are, respectively, the total energy and total entropy in the fluid at the beginning of the hydrodynamic evolution (see Sec.~\ref{s:initial}). In order to facilitate the comparison, the value of $E_0/S$ has been multiplied by a constant in each panel so that the mean matches that of the $\langle p_t\rangle$ distribution. 
    (a) 5-10\% centrality. (b) 50-55\% centrality. 
    }
    \label{fig:hydro}
\end{figure*}

A quantitative measure of the skewness of a random variable $x$ is the third centered moment, $\langle (x-\langle x\rangle)^3\rangle$, where angular brackets denote an average value with respect to the probability distribution of $x$.
It is usually positive when the tail is larger to the right than to the left, as in Figs.~\ref{fig:star} and \ref{fig:hydro}. 
The skewness is the third term in a systematic cumulant expansion, whose first and second terms are the mean and the variance, respectively. 

\subsection{Experimental analysis}
\label{s:experiment}

We first recall how the mean value of the $p_t$ distribution in a centrality class, which we denote by $\langle\!\langle p_t\rangle\!\rangle$, is evaluated in heavy-ion experiments. 
There are two ways of defining it, depending on whether one first averages over particles in an event~\cite{Adams:2005ka}, and then over all events, or whether one does both averages simultaneously~\cite{Abelev:2014ckr}. 
Specifically, the STAR collaboration defines~\cite{Adams:2005ka}:
\begin{equation}
\label{meanptstar}
\langle\!\langle p_t\rangle\!\rangle_{\rm STAR}\equiv\left\langle\frac{\sum_{i=1}^{N_{\rm ch}} p_{i}}{N_{\rm ch}}\right\rangle_{\rm ev},
\end{equation}
where $N_{\rm ch}$ denotes the number of charged particles in an event, $p_i$ is the transverse momentum of the $i$th particle, and angular brackets denote an average over events in a centrality class. 
On the other hand, the ALICE collaboration defines~\cite{Abelev:2014ckr}:
\begin{equation}
\label{meanptalice}
\langle\!\langle p_t\rangle\!\rangle_{\rm ALICE}\equiv\frac{\left\langle\sum_{i=1}^{N_{\rm ch}} p_{i}\right\rangle_{\rm ev}}{\left\langle N_{\rm ch}\right\rangle_{\rm ev}}.
\end{equation}
These definitions are almost equivalent, but not strictly equivalent when the multiplicity $N_{\rm ch}$ fluctuates event to event. 

Either convention can be used when analyzing the variance of dynamical $p_t$ fluctuations. We denote this variance by  $\left\langle \Delta p_{i} \Delta p_{j}\right\rangle$, where the subscripts $i,j$ are meant to remind that it is constructed from pair correlations, with $i\not= j$. 
The STAR collaboration defines it as~\cite{Adams:2005ka}:
\begin{equation}
\label{variancestar}
\left\langle \Delta p_{i} \Delta p_{j}\right\rangle_{\rm STAR}\equiv \left\langle \frac{\sum_{i,j\not =i} \left( p_{i}-\langle\!\langle p_t\rangle\!\rangle\right)\left( p_{j}-\langle\!\langle p_t\rangle\!\rangle\right)}{N_{\rm ch}\left(N_{\rm ch}-1\right)}   \right\rangle_{\rm ev}, 
\end{equation}  
where $\langle\!\langle p_t\rangle\!\rangle$ is defined by Eq.~(\ref{meanptstar}),
while the ALICE collaboration defines it as~\cite{Abelev:2014ckr}:
\begin{equation}
\label{variancealice}
\left\langle \Delta p_{i} \Delta p_{j}\right\rangle_{\rm ALICE}\equiv \frac{\left\langle\sum_{i,j\not =i} \left( p_{i}-\langle\!\langle p_t\rangle\!\rangle\right)\left( p_{j}-\langle\!\langle p_t\rangle\!\rangle\right)\right\rangle_{\rm ev}}{\left\langle N_{\rm ch}\left(N_{\rm ch}-1\right)\right\rangle_{\rm ev}}, 
\end{equation}  
where $\langle\!\langle p_t\rangle\!\rangle$ is defined by Eq.~(\ref{meanptalice}). 
Note that even though Eqs.~(\ref{variancestar}) and (\ref{variancealice}) involve double sums over $i$ and $j$, they can be expressed in terms of simple sums, which are much faster to compute. 
The corresponding formulas for Eq.~(\ref{variancestar}) are derived in Appendix~\ref{s:cumulants}. 
The skewness is the third centered moment, which we denote by $\left\langle \Delta p_{i} \Delta p_{j}\Delta p_{k}\right\rangle$. It is defined by straightforward generalizations of Eqs.~(\ref{variancestar}) and (\ref{variancealice}): 
\begin{widetext}
\begin{equation}
\label{skewnessstar}
\left\langle \Delta p_{i} \Delta p_{j} \Delta p_{k}\right\rangle_{\rm STAR}\equiv \left\langle \frac{\sum_{i,j\not =i,k\not=i,j} \left( p_{i}-\langle\!\langle p_t\rangle\!\rangle\right)\left( p_{j}-\langle\!\langle p_t\rangle\!\rangle\right)\left( p_{k}-\langle\!\langle p_t\rangle\!\rangle\right)}{N_{\rm ch}\left(N_{\rm ch}-1\right)\left(N_{\rm ch}-2\right)}   \right\rangle_{\rm ev}, 
\end{equation}  
where $\langle\!\langle p_t\rangle\!\rangle$ is defined by Eq.~(\ref{meanptstar}), and 
\begin{equation}
\label{skewnessalice}
\left\langle \Delta p_{i} \Delta p_{j} \Delta p_{k} \right\rangle_{\rm ALICE}\equiv \frac{\left\langle\sum_{i,j\not =i,k\not=i,j} \left( p_{i}-\langle\!\langle p_t\rangle\!\rangle\right)\left( p_{j}-\langle\!\langle p_t\rangle\!\rangle\right)\left( p_{k}-\langle\!\langle p_t\rangle\!\rangle\right)\right\rangle_{\rm ev}}{\left\langle N_{\rm ch}\left(N_{\rm ch}-1\right)\left(N_{\rm ch}-2\right)\right\rangle_{\rm ev}},
\end{equation}  
\end{widetext}
where $\langle\!\langle p_t\rangle\!\rangle$ is defined by Eq.~(\ref{meanptalice}). 
An efficient way of computing Eq.~(\ref{skewnessstar}) is detailed in Appendix~\ref{s:cumulants}.  Note that the ATLAS collaboration follows the same convention as the STAR collaboration in its recent analysis of transverse momentum fluctuations~\cite{Aad:2019fgl}. 

\subsection{Dimensionless observables}
\label{s:dimensionless}

We now define two dimensionless measures of the skewness, which should have less sensitivity to analysis details, in particular the acceptance in $p_t$, which varies depending on the detector. The first measure is the standardized skewness, defined by:
\begin{equation}
\label{standardized}
\gamma_{p_t}\equiv\frac{\left\langle \Delta p_{i} \Delta p_{j}\Delta p_{k} \right\rangle}{\left\langle \Delta p_{i} \Delta p_{j} \right\rangle^{3/2}}.
\end{equation}
This is a dimensionless quantity, but one expects it to depend on centrality and system size, as measured by the number of participant nucleons, $N_{\rm part}$.
In order to get an idea of this centrality dependence, let us assume for simplicity that dynamical fluctuations are proportional to statistical fluctuations. 
Statistical fluctuations are generated by the finite multiplicity, which is roughly proportional to $N_{\rm part}$. 
Therefore, the variance is proportional to $1/N_{\rm part}$, and the skewness to $1/N_{\rm part}^2$~\cite{Bhalerao:2019fzp}. 
Hence, one expects the standardized skewness to be  proportional to $1/\sqrt{N_{\rm part}}$.\footnote{These scaling rules are verified in a toy model in Appendix~\ref{s:sources}.}
The fact that it decreases with $N_{\rm part}$ is a consequence of the central limit theorem, which states that fluctuations are more Gaussian for a large system. 
Even though there is a priori no argument why dynamical fluctuations should be proportional to statistical fluctuations, it is reasonable to expect that the qualitative trends are similar, and that the standardized skewness is smaller in central collisions than in peripheral collisions. 
This can be seen by eye by comparing the full curves in Fig.~\ref{fig:hydro}(a) and Fig.~\ref{fig:hydro}(b). 
The corresponding values of the standardized skewness are 
$\gamma_{p_t}=0.26\pm 0.05$ and 
$\gamma_{p_t}=0.89\pm 0.08$ respectively, for these two centrality intervals. 

In order to eliminate the trivial dependence on the global size, we introduce a second measure of the skewness, which we dub the {\it intensive} skewness, and denote by $\Gamma$:
\begin{equation}
\label{intensive}
\Gamma_{p_t}\equiv\frac{\left\langle \Delta p_{i} \Delta p_{j}\Delta p_{k} \right\rangle\langle\!\langle p_t\rangle\!\rangle}{\left\langle \Delta p_{i} \Delta p_{j} \right\rangle^{2}}.
\end{equation}
With the above scaling rules, $\Gamma_{p_t}$ is {\it independent\/} of $N_{\rm part}$. 
In general, one does not expect $\Gamma_{p_t}$ to be independent of $N_{\rm part}$, but its centrality dependence should be milder than that of the standardized skewness. 
This is confirmed by an explicit calculation for the results shown as full lines in Fig.~\ref{fig:hydro}(a) and Fig.~\ref{fig:hydro}(b), which gives  $\Gamma_{p_t}=7.5\pm 1.5$ and $\Gamma_{p_t}=10.6\pm 0.9$ (see also Sec.~\ref{s:hydro} and Fig.~\ref{fig:hydroskew}). 

\subsection{Baseline for the intensive skewness}
\label{s:baseline}

We have pointed out in Sec.~\ref{s:evidence} that a positive skewness is anyway expected as a result of the positiveness of $p_t$. 
A natural baseline is provided by the distribution of $\langle p_t\rangle$ for mixed events, which are constructed by mixing random particles from different events. 
Since mixed events are made of $N$ independent particles, the cumulants of the distribution of $\sum_{i=1}^N p_{t,i}$ are the cumulants of the distribution of $p_t$ for a single particle, multiplied by $N$. 
The dependence on $N$ cancels in the intensive skewness (\ref{intensive}). 
Therefore, the intensive skewness for mixed events reduces to that for a single particle:
\begin{equation}
\label{independent}
\Gamma_{{\rm independent}}\equiv\frac{\left\langle (p_t-\langle p_t\rangle)^3 \right\rangle\langle p_t\rangle}{\left\langle (p_t-\langle p_t\rangle)^2\right\rangle^{2}},
\end{equation}
where angular brackets denote an average over $p_t$ with the weight $dN/dp_t$. 
The value of $\Gamma_{{\rm independent}}$ can easily be evaluated using available data on $dN/dp_t$. 
It is typically around $3$, and increases mildly as a function of centrality percentile. 
Specifically, it varies between $2.8$ and $3.0$ in Au+Au collisions at $\sqrt{s_{\rm NN}}=200$~GeV~\cite{Back:2003qr} and between $3.2$ and $3.7$ in Pb+Pb collisions at $\sqrt{s_{\rm NN}}=5.02$~TeV~\cite{Acharya:2018qsh} (see Fig.~\ref{fig:hydroskew}). 

\section{Results from hydrodynamic simulations}
\label{s:hydro}
\begin{figure*}[ht]
    \centering
    \includegraphics[width=.95\linewidth]{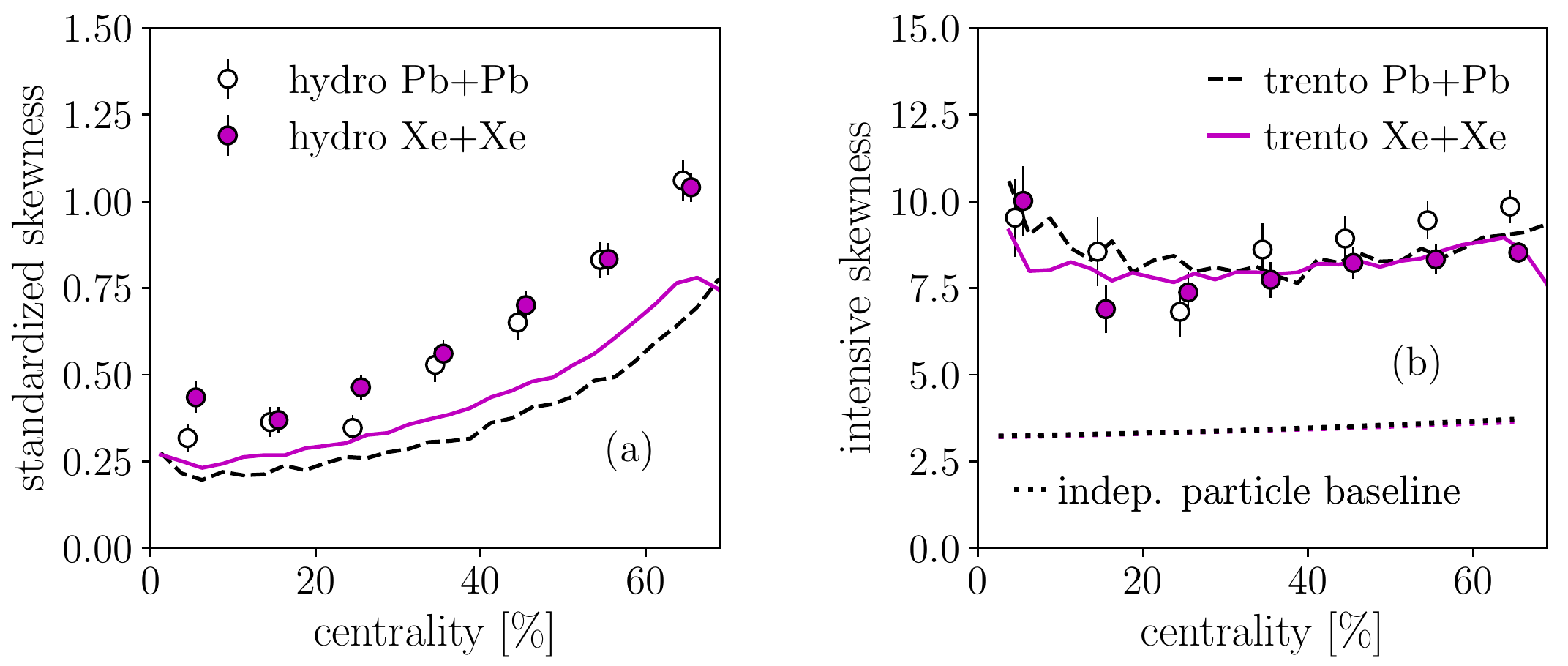}
    \caption{(Color online)
    Results of our hydrodynamic calculations for:
    (a) the standardized skewness, $\gamma_{p_t}$, defined by Eq.~(\ref{standardized}); (b) the intensive skewness, $\Gamma_{p_t}$, defined by Eq.~(\ref{intensive}), in Pb+Pb collisions at $\sqrt{s_{\rm NN}}=5.02$~TeV (open symbols) and Xe-Xe collisions at $\sqrt{s_{\rm NN}}=5.44$~TeV (full symbols), as a function of the centrality percentile. Error bars represent the statistical error, due to the finite number of hydrodynamic events, estimated via jackknife resampling. 
    The open and full symbols have been slightly shifted to the left and to the right, respectively, for the sake of readability. Lines are the same quantities as symbols, where one replaces $\langle p_t\rangle$ with the value of $E_0/S$ at the beginning of the hydrodynamic evolution (Sec.~\ref{s:initial}). The dotted line in panel (b) represents the baseline defined by Eq.~(\ref{independent}) for the intensive skewness, evaluated using the measured $p_t$ spectra~\cite{Acharya:2018qsh}. 
  }
    \label{fig:hydroskew}
\end{figure*}

Evaluating the skewness in event-by-event hydrodynamics is much simpler than in experiment because, as explained in Sec.~\ref{s:evidence}, one need not worry about statistical fluctuations. 
One evaluates $\langle p_t\rangle$ for each initial condition by integrating the continuous momentum distribution resulting from the hydrodynamic expansion. 
The mean transverse momentum in a centrality class, $\langle\!\langle p_t\rangle\!\rangle$, is obtained by averaging $\langle p_t\rangle$ over initial conditions. The variance and the skewness are then defined by:
\begin{eqnarray}
\label{hydrocumulants}
\left\langle \Delta p_{i} \Delta p_{j} \right\rangle_{\rm hydro}&=& \left\langle \left(\langle p_t\rangle -\langle\!\langle p_t\rangle\!\rangle\right)^2\right\rangle_{\rm ev}\cr
\left\langle \Delta p_{i} \Delta p_{j} \Delta p_{k} \right\rangle_{\rm hydro}&=&\left\langle \left(\langle p_t\rangle -\langle\!\langle p_t\rangle\!\rangle\right)^3\right\rangle_{\rm ev},
\end{eqnarray}
where the outer angular brackets denote an average over initial conditions.  

Figure~\ref{fig:hydroskew} presents our result for the standardized skewness [panel (a)], and the intensive skewness [panel (b)] for Xe+Xe and Pb+Pb collisions, as a function of the centrality percentile, using the same hydrodynamic calculation as in Sec.~\ref{s:evidence}. 
The standardized skewness in panel (a) increases as a function of the centrality percentile, as already observed in Fig.~\ref{fig:hydro}, reflecting the fact that larger centrality implies a smaller number of participant nucleons. One also expects the standardized skewness to be larger in the smaller system, Xe+Xe, although, within our numerical precision, this is not observed in all the centrality bins. 
 Since, as discussed in Sec.~\ref{s:evidence}, our hydrodynamic model overestimates $\langle p_t\rangle$ fluctuations, it is likely to also overestimate the standardized skewness, and the results  in Fig.~\ref{fig:hydroskew} (a)  should not be considered a quantitative prediction. 
 
The intensive skewness should be more robust against a rescaling of the fluctuations. 
We therefore hope that our hydrodynamic calculation, even though they overestimate $\langle p_t\rangle$ fluctuations, have some predictive power for this quantity, shown in panel (b).  
It depends mildly on the collision species and the collision centrality. 
While the baseline defined by Eq.~(\ref{independent}) is between 3 and 4, the prediction from hydrodynamics is much larger, $7<\Gamma_{p_t} <10$. 
The prediction that the skewness is ``larger than trivial'' is our main point.

\section{Origin of the skewness}
\label{s:initial}

We now investigate the origin of the large positive skewness of $\langle p_t\rangle$
fluctuations found in hydrodynamic calculations. As it was shown in Refs.~\cite{Gardim:2020sma,Giacalone:2020dln}, if one looks at events with the same \textit{initial} entropy (which experimentally can be achieved to a good approximation by fixing the final-state multiplicity), then $\langle p_t\rangle$ is tightly correlated with the total energy of the fluid at the beginning of the hydrodynamic evolution, $E_0$. Intuitively, this is due to the fact that the momentum is a function of the energy, and thus, if the number of particles is fixed, it is the energy that determines the mean transverse momentum.
The nontrivial aspect of this correspondence is that the correlation of $\langle p_t\rangle$ is tighter with the initial energy, $E_0$, than with the energy at freeze-out~\cite{Gardim:2020sma}, even though particles are emitted at freeze-out. The goal of this section is to show that, at fixed centrality, one expects the skewness of $\bra p_t \ket$ fluctuations to be driven by the skewness of $E_0$ fluctuations.

Although the relation between $\langle p_t\rangle$ and $E_0$ is not quite linear, we can relate their fluctuations in a simplified, \textit{effective} hydrodynamic description~\cite{Gardim:2019xjs}. This description replaces the space-time evolution of the quark-gluon plasma with an equivalent uniform gas at an effective temperature, $T_{\rm eff}$, that contains the same total entropy and total energy as the quark-gluon plasma at freezeout.
In this effective description, the final-state $\langle p_t\rangle$ is proportional to $T_{\rm eff}$, whereas $E_0$ is proportional to $\epsilon/s$, where the energy density, $\epsilon$, and the entropy density, $s$, are evaluated at temperature $T_{\rm eff}$. 
The fluctuations of $E_0$ and those of $\bra p_t \ket$ can be then related through the equation of state. 

Let us first derive a relation between the relative variation of $\langle p_t\rangle$ and that of $E_0$ in the regime of small fluctuations. First note that the relative $\bra p_t \ket$ variation is related to that of the effective entropy density, $s_{\rm eff}$, through:
 \begin{equation}
 \label{ptvar}
 d\ln\langle p_t\rangle=d\ln T_{\rm eff}=c_{s,\rm eff}^2 d\ln s_{\rm eff},
 \end{equation}
 where $c_{s,\rm eff}=(d\ln T/d\ln s)^{1/2}$ is the speed of sound at temperature $T_{\rm eff}$. Similarly, the relative variation of $E_0$ is given by:
 \begin{align}
 \label{eivar}
\nonumber  d\ln E_0&=d\ln\left(\frac{\epsilon_{\rm eff}}{s_{\rm eff}}\right)
 = \frac{d\epsilon_{\rm eff}}{\epsilon_{\rm eff}}-\frac{ds_{\rm eff}}{s_{\rm eff}} \\
\nonumber  =& \frac{T_{\rm eff}ds_{\rm eff}}{\epsilon_{\rm eff}}-\frac{ds_{\rm eff}}{s_{\rm eff}}
 = \frac{(\epsilon_{\rm eff}+P_{\rm eff}) ds_{\rm eff}-\epsilon_{\rm eff} ds_{\rm eff}}{\epsilon_{\rm eff} s_{\rm eff}} \\
  =& (P_{\rm eff}/\epsilon_{\rm eff})d\ln s_{\rm eff}, 
 \end{align}
where we have used the thermodynamic identities $d\epsilon=Tds$ and $\epsilon+P=Ts$. 
Combining the last two equations, one predicts 
\begin{equation}
\label{ratiosigma}
\frac{\sigma(\langle p_t\rangle)}{\langle\!\langle p_t\rangle\!\rangle}=c_s^2\frac{\epsilon_{\rm eff}}{P_{\rm eff}}\frac{\sigma(E_0)}{\langle E_0\rangle}\simeq 1.24\frac{\sigma(E_0)}{\langle E_0\rangle}
\end{equation}
where $\sigma(\langle p_t\rangle)$ and $\sigma(E_0)$ denote, respectively, the standard deviation of $\langle p_t\rangle$ and $E_0$, and, in the last equality, we have used $T_{\rm eff}=222$~MeV for 5.02 TeV Pb+Pb collisions~\cite{Gardim:2019xjs}, at which we have evaluated the thermodynamic quantities using the lattice QCD equation of state~\cite{Borsanyi:2013bia}. Note that the relative fluctuations of $E_0$ is equal to that of $E_0/S$, where $S$ is the total entropy, when $S$ is kept fixed. To correct for potential effects of finite-sized centrality intervals, or entropy production due to viscosity, one should replace $E_0$ by $E_0/S$ in Eq.~(\ref{ratiosigma}). 

To prove that Eq.~(\ref{ratiosigma}) provides a meaningful prediction, we show in Fig.~\ref{fig:hydro} the distribution of $E_0/S$, which we have rescaled so that its mean value coincides with that of $\langle p_t\rangle$. 
Equation~(\ref{ratiosigma}) predicts that the fluctuations of $\bra p_t \ket$ are larger than those of the rescaled $E_0/S$ by a factor $1.24$. 
One sees by eye on the figure that the distribution of $\langle p_t\rangle$ is broader. 
The factor is somewhat larger than $1.24$ (it is $1.3$ in panel (a) and $1.6$ in panel (b)), and we do not understand the origin of this difference. 

Let us now move on to the skewness. 
Even if $\langle p_t\rangle$ is exactly determined by the initial energy $E_0$ on an event-by-event basis, the skewness of $\langle p_t\rangle$ is {\it not\/} trivially related to the skewness of $E_0$. 

To see this, let us consider a background-fluctuation splitting, $\langle p_t\rangle=\langle\!\langle p_t\rangle\!\rangle+\delta p_t$, $E_0=\langle E_0\rangle+\delta E_0$, and assume that the average transverse momentum is a generic function of the initial energy, $\langle p_t\rangle =f(E_0)$. To leading order, the expansion of $f(E_0)$ in powers of the fluctuation leads to: 
\begin{align}
\label{e0expansion}
\nonumber \langle (\delta p_t)^2\rangle =& f'(\langle E_0\rangle)^2 \langle (\delta E_0)^2\rangle \\
\nonumber \langle (\delta p_t)^3\rangle =& f'(\langle E_0\rangle)^3 \langle(\delta E_0)^3\rangle \\
&+\frac{3}{2} f'(\langle E_0\rangle)^2 f^{\prime\prime}(\langle E_0\rangle)\left(  \langle(\delta E_0)^4\rangle- \langle(\delta E_0)^2\rangle^2\right).
\end{align}
In general, the two terms in the second equation are of the same order of 
magnitude.\footnote{This is estimated from the fact that both $ \bra E_0 \ket \langle(\delta E_0)^3\rangle$ and $\langle(\delta E_0)^4\rangle-\langle(\delta E_0)^2\rangle^2$ are of order $\bra (\delta E_0)^2 \ket^2$. As a consequence, the magnitude of the second term relative to the first  is $\sim E_0 f^{\prime\prime}(E_0)/f'(E_0)$, evaluated at $\langle E_0\rangle$. This is the relative variation of $f'(E_0)$ over a range of order $E_0$, which is typically of order unity.}
The function $f(E_0)$ represents, however, the variation of the temperature of the system as a function of the energy over entropy ratio, $\epsilon/s$. This variation is nearly linear. One can thus neglect the term proportional to $f^{\prime\prime}(\langle E_0\rangle)$ in Eq.~(\ref{e0expansion}). By doing so, one immediately sees that the standardized skewness of $\bra p_t \ket$ fluctuations, $\gamma_{p_t}=\bra (\delta p_t)^3 \ket /\bra (\delta p_t)^2 \ket^{3/2}$, is the same as that of $E_0$ fluctuations:
\begin{equation}
\label{sskeweff}
\gamma_{p_t}\simeq \gamma_{E_0}.
\end{equation}
On the other hand, the intensive skewness is different: 
\begin{eqnarray}
\label{iskeweff}
\Gamma_{p_t} \equiv  \frac{\bra (\delta p_t)^3 \ket}{\bra (\delta p_t)^2 \ket / \bra\!\bra p_t \ket\!\ket} &\simeq& \frac{\langle\!\langle p_t\rangle\!\rangle}{\langle E_0\rangle f'(\langle E_0\rangle)}\Gamma_{E_0}\cr
&\simeq& \frac{\langle\!\langle p_t\rangle\!\rangle}{\langle E_0\rangle} \frac{\sigma_{p_t}}{\sigma_{E_0}}\Gamma_{E_0}\cr
&\simeq& \frac{P_{\rm eff}}{\epsilon_{\rm eff}}\frac{1}{c_{s,\rm eff}^2}\Gamma_{E_0}\cr
&\simeq&0.8\,\Gamma_{E_0},
\end{eqnarray}
where we have used Eq.~(\ref{ratiosigma}). 
Thus, even though it is not obvious why the skewnesses of $\bra p_t \ket$ fluctuations and $E_0$ fluctuations should be closely related, they are in practice if one can neglect $f^{\prime\prime}(E_0)$ in Eq.~(\ref{e0expansion}). 

Figure~\ref{fig:hydroskew} displays the standardized skewness and the intensive skewness of the distribution of $E_0/S$
(we swap $E_0$ for $E_0/S$ to account for the fluctuations of centrality within the bin, as explained above).
One sees that  $\gamma_{E_0}$ is smaller than $\gamma_{p_t}$,\footnote{Note that $\gamma_{E_0}$ is larger in Xe+Xe collisions than in Pb+Pb collisions, as expected from the smaller system size.} while $\Gamma_{E_0}$ is comparable to $\Gamma_{p_t}$. 
The fact that Eq.~(\ref{sskeweff}) is not precisely verified seems to imply that the correction from the last line of Eq.~(\ref{e0expansion}) is not negligible. 
However, the main features displayed by the skewness of $\langle p_t\rangle$ fluctuations stem from the skewness of the initial energy, $E_0$. 
In particular, the large intensive skewness of $\langle p_t\rangle$ fluctuations, which is our main prediction, stems from that of $E_0$ fluctuations. 
Note that a more quantitative understanding may be achieved by improving the initial-state predictor. In a recent preprint Schenke, Shen and Teaney~\cite{Schenke:2020uqq} studied the goodness of various estimators of $\langle p_t\rangle$, and found that an improved predictor, especially for peripheral collisions, can be obtained by adding a dependence on the elliptical area of the system~\cite{Bozek:2017elk}. We do not investigate this possibility here.

\section{Relating the skewness to initial density fluctuations}
\label{s:perturbative}

The results of the previous section show that the skewness of $\bra p_t \ket$ fluctuations originates from the skewness of $E_0$ fluctuations. The fact that the latter skewness is positive, though, is specific to the model used in the numerical evaluation, i.e., a \trento{} parametrization tuned to reproduce some sets of experimental data. 
In this section, we argue that the prediction that $\langle p_t\rangle$ fluctuations have positive skewness is more general, and does not rely on a specific model of initial conditions. For this purpose, we derive formulas for the variance and the skewness of $E_0$ fluctuations for a generic fluctuating initial density profile. 

\subsection{Formalism}

Our study is limited to boost-invariant ideal hydrodynamics for simplicity, and neglects initial transverse flow~\cite{Vredevoogd:2008id,Kurkela:2018wud}.
The hydrodynamic evolution is then determined by the entropy density field at the initial condition, $s(x)$, where $x$ denotes a point in the transverse plane. 
We consider an ensemble of events with the same geometry (same positions of incoming nuclei) and same total entropy, $\int s(x) dx$. The fluctuations of the field $s(x)$ within this ensemble of events can be characterized by its $n$-point correlation functions. We assume that, for any event, $s(x)$ can be decomposed as a fluctuation on top of a background: $s(x)=\langle s(x)\rangle+\delta s(x)$, where $\langle s(x)\rangle$, or 1-point function, is the average value of $s(x)$ for a fixed $x$, and $\delta s(x)$ is the fluctuation. 
Observables are evaluated through a perturbative expansion in powers of the fluctuation. This approach is identical to that of Refs.~\cite{Blaizot:2014nia,Gronqvist:2016hym,Bhalerao:2019uzw,Bhalerao:2019fzp}. 

The only technical difference with these references is that we take now $s(x)$ as the fundamental field instead of the energy density, $\epsilon(x)$. 
This choice simplifies the algebra, because the centrality is defined in terms of the total entropy, not energy. 
At a fixed centrality, the total entropy is fixed, which implies:
\begin{equation}
\label{sumrule}
\int_x \delta s(x)=0,
\end{equation}
where we use the shortcut $\int_x$ for the integration over the transverse plane, which is a double integral. 

Initial-state fluctuations are characterized by the statistical properties of the field $\delta s(x)$ or, equivalently, by its $n$-point functions. 
The connected $2$-point function is the average over events of $\delta s(x_1)\delta s(x_2)$. It characterizes how fluctuations at different points $x_1$ and $x_2$ are correlated with one another. 
We assume that all fluctuations are local, which implies that correlations are short ranged. 
Under this condition, one can write the two-point function in the form~\cite{Gronqvist:2016hym}:
\begin{equation}
\label{2pointc}
\langle\delta s(x_1)\delta s(x_2)\rangle=\kappa_2(x_1)\delta(x_1-x_2)- 
\frac{\kappa_2(x_1)\kappa_2(x_2)}{\int_x\kappa_2(x)},
\end{equation}
where we assimilate the short range correlation to a Dirac peak, $\delta(x_1-x_2)$, with a positive $x$-dependent amplitude, $\kappa_2(x)$, which represents the density of variance of the entropy field. 
Equation~(\ref{sumrule}) implies that the two-point function must vanish upon integration over $x_1$ or $x_2$. 
This is guaranteed by the last term in the right-hand side of Eq.~(\ref{2pointc}).

To evaluate the skewness, we shall also need the three-point function of the density field. 
As shown in Ref.~\cite{Gronqvist:2016hym}, for short-range correlations the three-point function at fixed total entropy can be written in the form:
\begin{align}
\label{3pointc}
\nonumber \langle\delta s(x_1)\delta s(x_2)\delta s(x_3)\rangle&=\kappa_3(x_1)\delta(x_1-x_2)\delta(x_1-x_3)\\
\nonumber&-\frac{\kappa_3(x_1)\delta(x_1-x_2)\kappa_2(x_3)+{\rm
  perm.}}{\int_x\kappa_2(x)}\\ 
\nonumber &+\frac{\kappa_3(x_1)\kappa_2(x_2)\kappa_2(x_3)+{\rm
  perm.}}{\left(\int_x\kappa_2(x)\right)^2}\\
 &-\frac{\int_x\kappa_3}{\left(\int_x\kappa_2\right)^3}\kappa_2(x_1)\kappa_2(x_2)\kappa_2(x_3),
\end{align}
where the second and third lines must be summed over circular permutations of $x_1$, $x_2$, $x_3$.
The first term in the right-hand side is the contribution of the short-range correlation, and $\kappa_3(x)$ is the ``density of skewness'', in the same way as $\kappa_2(x)$ is the density of variance. 
Note that $\kappa_3(x)$ is typically positive everywhere (e.g. for Poisson fluctuations), even though this is not a mathematical requirement.
The additional terms in Eq.~(\ref{3pointc}) are contributions from the condition that all events have the same total entropy. 
This expression is consistent with the sum rule (\ref{sumrule}), as can be checked upon integration over $x_1$ (or $x_2$ or $x_3$, by symmetry). 
Note that the three-point function involves both $\kappa_2(x)$ and $\kappa_3(x)$, and it is linear in $\kappa_3(x)$.

\subsection{Variance of initial energy fluctuations}
\label{s:var}

Equipped with this formalism, we evaluate the fluctuations of the initial energy, $E_0$. 
This quantity is given by the integral of the energy density, $\epsilon(x)$, which is related to $s(x)$ through the equation of state:
\begin{equation}
  \label{defE}
E_0=\int_x \epsilon\left (s(x) \right). 
\end{equation}
We then write $s(x)=\langle s(x)\rangle+\delta s(x)$, and expand in powers of $\delta s(x)$.
To first order in $\delta s(x)$, one can write $E_0=\langle E_0\rangle+\delta E_0$, with
\begin{align}
\label{E1}
\nonumber  \langle E_0\rangle &=\int_x \epsilon(\langle s(x)\rangle), \\
\delta E_0&=\int_x T(x)\delta s(x),  
\end{align}
where $T(x)$ is the temperature corresponding to the average entropy density, $\langle s(x)\rangle$, and we have used the thermodynamic identity $d\epsilon=Tds$. 
The variance of the energy is:
\begin{equation}
  \label{varE}
  \langle \delta E_0^2\rangle
  =\int_{x_1,x_2}T(x_1)T(x_2)\langle\delta s(x_1)\delta s(x_2)\rangle.
\end{equation}
Using the expression (\ref{2pointc}) of the two-point function, one obtains
\begin{equation}
\label{varE2}
\langle \delta E_0^2\rangle
=\int_x T(x)^2\kappa_2(x) -\frac{\left(\int_x T(x)\kappa_2(x)\right)^2}{\int_x\kappa_2(x)},
\end{equation}
where the last term in the right-hand side comes from the condition that the total entropy is fixed. 
This equation can be rewritten in a simpler form by introducing the average temperature $\bar T$ defined by:  
\begin{equation}
  \label{defbarT}
  \bar T\equiv \frac{\int_x T(x)\kappa_2(x)}{\int_x\kappa_2(x)}.
\end{equation}
It is the temperature averaged over the transverse plane, weighted with the variance of the entropy field $\kappa_2(x)$.   
With this notation, Eq.~(\ref{varE2}) can be rewritten as
\begin{equation}
  \label{varE3}
  \langle \delta E_0^2\rangle
=\int_x (T(x)-\bar T)^2\kappa_2(x). 
\end{equation}
Note that the condition that all events have the same total entropy results in the substitution $T(x)\to T(x)-\bar T$. 

Let us comment on the physical implication of Eq.~(\ref{varE3}). 
In this equation, $T(x)$ denotes the temperature profile at the beginning of the hydrodynamic expansion, that is, when the temperature is the highest, and $\bar T$ its value averaged over $x$. 
The difference $T(x)-\bar T$ is a temperature difference, which is proportional to $c_s^2$. 
Therefore, one expects the relative fluctuation of $E_0$ to be itself proportional to $c_s^2$, where $c_s$ is the velocity of sound at the beginning of the hydrodynamic calculation. 
This is checked by an explicit calculation in Appendix~\ref{s:cs}. 
This correspondence only holds in ideal hydrodynamics, and viscous corrections are large at early times, therefore, its relevance to the phenomenology is questionable. 
However, it suggests that the physics of $\langle p_t\rangle$ fluctuations might open a window onto early-time thermodynamics.

\subsection{Skewness of initial energy fluctuations}

We now evaluate the skewness of the distribution of $E_0$. This is a higher-order quantity, therefore, we need to expand the energy density to order 2 in $\delta s$:
\begin{equation}
\label{order2}
  \epsilon(s(x))=\epsilon(\langle s(x)\rangle)+T(x)\delta s(x)+\frac{1}{2}T'(x) \delta s(x)^2,
\end{equation}
where we define
\begin{equation}
    T'(x)\equiv\frac{dT}{ds}=c_s^2(x) \frac{T(x)}{\langle s(x)\rangle},
\end{equation}
where $c_s(x)$ is the speed of sound at the temperature $T(x)$. 
With the second order term taken into account, Eq.~(\ref{E1}) is replaced by:
\begin{align}
\label{E2}
\nonumber \langle E_0\rangle &=\int_x \epsilon(\langle s(x)\rangle)+\frac{1}{2}\int_xT'(x) \langle\delta s(x)^2\rangle, \\
\delta E_0&=\int_x T(x)\delta s(x)+\frac{1}{2}\int_x T'(x) \left(\delta s(x)^2-\langle\delta s(x)^2\rangle\right). 
\end{align}
The skewness is the third centered moment, that is, $\langle \delta E_0^3\rangle$. 
To leading order in the fluctuations, one must keep all terms of order 3 and 4 in $\delta s$, which contribute to the same order after averaging over events.
We write
\begin{equation}
\label{deltae3dec}
  \langle \delta E_0^3\rangle=\langle \delta E_0^3\rangle_3+\langle \delta E_0^3\rangle_4,
\end{equation}
where we separate the contributions of terms of order $\delta s^3$ and $\delta s^4$.

The contribution of order $\delta s^3$ is obtained by keeping only the first term in the second line of Eq.~(\ref{E2}):
\begin{equation}
\label{deltaE3}
  \langle \delta E_0^3\rangle_3=\int_{x_1,x_2,x_3}T(x_1)T(x_2)T(x_3)\langle
  \delta s(x_1)\delta s(x_2)\delta s(x_3)\rangle.
\end{equation}
It involves the three-point function of the density field. 
Inserting Eq.~(\ref{3pointc}) into Eq.~(\ref{deltaE3}), one obtains, after some algebra, a compact result: 
\begin{equation}
  \label{deltae33}
\langle \delta E_0^3\rangle_3=\int_x \left(T(x)-\bar T\right)^3\kappa_3(x), 
\end{equation}
where $\bar T$ is defined by Eq.~(\ref{defbarT}).
As in Eq.~(\ref{varE3}), the condition that all events have the same entropy results in the substitution $T(x)\to T(x)-\bar T$.

We finally evaluate $\langle\delta E^3\rangle_4$, which is the contribution obtained by expanding two factors of $\delta E$ to order $\delta s$ and the third factor to order $\delta s^2$.
One is led to evaluate the average value of quantities such as:
\begin{equation}
A(x_1,x_2,x_3)\equiv\delta s(x_1)\delta s(x_2)\left(\delta s(x_3)^2-\langle \delta s(x_3)^2\rangle\right),
\end{equation}
where four-point averages can be computed using Wick's theorem, which gives:
\begin{equation}
\label{wick}
  \langle A(x_1,x_2,x_3)\rangle
  =2\langle \delta s(x_1)\delta s(x_3)\rangle\langle \delta s(x_2)\delta s(x_3)\rangle,
\end{equation}
where the right-hand side involves the 2-point function, Eq.~(\ref{2pointc}).
After some algebra, one obtains
\begin{equation}
  \label{deltae34}
\langle\delta E_0^3\rangle_4=3\int_x(T(x)-\bar T)^2T'(x)\kappa_2(x)^2. 
\end{equation}

The integrand is everywhere positive, so that $\langle\delta E_0^3\rangle_4$ is positive. It is interesting to note that the intermediate calculations involve the variance of the entropy density at a given point, i.e., the term $\langle\delta s(x)^2\rangle$ in Eq.~(\ref{E2}). 
This quantity is sensitive to the scale of inhomogeneities~\cite{Noronha-Hostler:2015coa}, that is, to the transverse size of the ``hot spots'' in the initial density profile.
However, this dependence cancels in Eq.~(\ref{wick}), and the final results depend only on the functions $\kappa_n(x)$, which are integrated over the relative distance.  
This implies that both the width of $\langle p_t\rangle$ fluctuations and their skewness should have limited sensitivity to short-range, subnucleonic fluctuations, in the same way as anisotropic flow fluctuations~\cite{Bhalerao:2011bp,Noronha-Hostler:2015coa,Mazeliauskas:2015vea,Gardim:2017ruc}.
They are on the other hand potentially useful probes of early-time thermodynamics, as suggested at the end of Sec.~\ref{s:var}. 

To conclude, let us write down our final formula for the skewness, Eq.~(\ref{deltae3dec}). It is the sum of the contributions (\ref{deltae33}) and (\ref{deltae34}):
\begin{align}
  \label{final}
\nonumber \langle \delta E_0^3\rangle &=  \int_x \left(T(x)-\bar T\right)^3\kappa_3(x) \\
& + 3\int_x(T(x)-\bar T)^2T'(x)\kappa_2(x)^2.
\end{align}
The second term is always positive, while the first contribution is typically negative, but smaller in magnitude.
In Appendix~\ref{s:sources} we check explicitly that, in the simple case of identical, localized sources with a Gaussian distribution, where all integrals can be carried out analytically, the second term indeed dominates over the first term so that the skewness is positive.
The contribution (\ref{deltae34}) provides, thus, a model-independent explanation for the positive skewness of $E_0$ fluctuations, and consequently of $\langle p_t\rangle$ fluctuations.

\section{Conclusions}
Hydrodynamics predicts that the event-by-event fluctuations of the mean transverse momentum, $\langle p_t\rangle$, have positive skew. This prediction could be verified straightforwardly in experiments, following the analysis procedures explained in this manuscript. 
The skewness is the simplest manifestation of non-Gaussian fluctuations.  
Non-Gaussian fluctuations are generic in small systems.  
Their study has proven useful in the context of anisotropic flow fluctuations in peripheral nucleus-nucleus collisions~\cite{Giacalone:2016eyu,Sirunyan:2017fts,Acharya:2018lmh} and in proton-nucleus collisions~\cite{Yan:2013laa,Sirunyan:2019pbr}. 
In the case of $\langle p_t\rangle$ fluctuations, the non-Gaussianity should be easy to measure all the way up to central nucleus-nucleus collisions. 

We have argued that $\langle p_t\rangle$ fluctuations result from fluctuations of the energy of the fluid when the hydrodynamic expansion starts. 
This confirms that dynamical $\langle p_t\rangle$ fluctuations are a collective effect, much in the same way as anisotropic flow. 
It also implies, more specifically, that they are sensitive to the early thermodynamics of the quark-gluon plasma, corresponding to the highest temperatures achieved in the collision.
This insight into high temperatures is very unique.
Other hadronic observables, such as the average transverse momentum, also provide insight about the thermodynamics, but at a much lower temperature  (around $T\sim 220$~MeV in 5.02 TeV Pb-Pb collisions~\cite{Gardim:2019xjs}).

The width of $\langle p_t\rangle$ fluctuations alone cannot constrain early thermodynamics, because it also depends on the model of initial conditions. 
However, we have shown that by measuring simultaneously the skewness and the width, one can combine them in such a way that the sensitivity to the initial condition model is significantly reduced.
We have introduced a dimensionless quantity, the intensive skewness, and shown in a simplified model (Appendix~\ref{s:sources}) that it only depends on the speed of sound at the time when the hydrodynamic expansion starts. 
Our simplified model assumes a constant speed of sound, and does not take into account that the relation between $\langle p_t\rangle$ and the initial energy is nonlinear. 
We have also carried out full hydrodynamic calculations, but using a model which overestimates the width of $\langle p_t\rangle$  fluctuations. 
This study is only preliminary, and more are needed. 
Based on the work done in this paper, we conjecture that the intensive skewness should lie between 7 and 10 in nucleus-nucleus collisions, about twice as large as the baseline from independent particles. 
We also predict that it depends little on the collision centrality and of the size of the colliding nuclei.

\section*{Acknowledgments}
We thank Chun Shen, Bjoern Schenke, and Derek Teaney for useful discussions. FGG was supported by CNPq (Conselho Nacional de Desenvolvimento Cientifico) grant 312932/2018-9, by  INCT-FNA grant 464898/2014-5 and FAPESP grant 2018/24720-6.
G.G., and J.-Y.O. were supported by USP-COFECUB (grant Uc Ph 160-16, 2015/13). J.N.H. acknowledges the support of the Alfred P. Sloan Foundation, support from the US-DOE Nuclear Science Grant No.  de-sc0019175.

\appendix
\section{Coding the skewness analysis}
\label{s:cumulants}

In this Appendix, we explain how to efficiently compute the skewness. 
We choose the first definition, Eq.~(\ref{skewnessstar}), but similar algebraic manipulations can be carried out to simplify the second definition, Eq.~(\ref{skewnessalice}). 
In every event, one evaluates the moments of the $p_t$ distributions, defined by 
\begin{equation}
\label{defqn}
Q_n=\sum_{i=1}^{N_{\rm ch}} (p_i)^n,
\end{equation} 
where $n=1,2,3$, $p_i$ is the transverse momentum of particle $i$, and the sum runs over all the charged particles detected in the event.
Sums over pairs and triplets of particles can be expressed simply in terms of these moments: 
\begin{align}
\nonumber \sum_{i,j\not =i} p_{i}p_{j}&=(Q_1)^2-Q_2 , \\
\sum_{i,j\not =i,k\not=i,j} p_{i}p_{j}p_{k}&=(Q_1)^3-3Q_2Q_1+2 Q_3.
\end{align}
These equations express the multiple sums in the left-hand side in terms of simple sums, which are faster to evaluate.
They are specific cases of Eqs.~(11) and (14) of Ref.~\cite{DiFrancesco:2016srj} in the case of a unique set of particles, 
$A_1=A_2=A_3$.

With these notations, Eqs.~(\ref{meanptstar}) and (\ref{variancestar}) can be rewritten in the form: 
\begin{align}
\nonumber \langle\!\langle p_t\rangle\!\rangle_{\rm STAR}&=\left\langle\frac{Q_1}{N_{\rm ch}}\right\rangle, \\
\nonumber \left\langle \Delta p_{i} \Delta p_{j}\right\rangle_{\rm STAR}&=\left\langle \frac{(Q_1)^2-Q_2}{N_{\rm ch}\left(N_{\rm ch}-1\right)}  \right\rangle
-\left\langle\frac{Q_1}{N_{\rm ch}}\right\rangle^2, \\
\end{align}  
where angular brackets denote an average value over events in a narrow centrality bin. 
Note that these expressions are strictly equivalent to those used by the STAR collaboration (Eqs.~(1)-(4) of Ref.~\cite{Adams:2005ka}), even though they are written in a different form. 

Finally, Eq.~(\ref{skewnessstar}) can be rewritten in the form: 
\begin{widetext}
\begin{align}
\nonumber \left\langle \Delta p_{i} \Delta p_{j} \Delta p_{k}\right\rangle_{\rm STAR}&=\left\langle \frac{(Q_1)^3-3Q_2Q_1+2 Q_3}{N_{\rm ch}\left(N_{\rm ch}-1\right)\left(N_{\rm ch}-2\right)} \right\rangle-3\left\langle \frac{(Q_1)^2-Q_2}{N_{\rm ch}\left(N_{\rm ch}-1\right)}  \right\rangle\left\langle\frac{Q_1}{N_{\rm ch}}\right\rangle+2\left\langle\frac{Q_1}{N_{\rm ch}}\right\rangle^3.
\end{align}  
\end{widetext}
This equation expresses the skewness in terms of the simple sums in Eq.~(\ref{defqn}), which are much faster to compute than the multiple sums in Eq.~(\ref{skewnessstar}). 
It has been advocated~\cite{Schenke:2020uqq} that the analysis of $\langle p_t\rangle$ fluctuations should be done by enforcing rapidity gaps between the particles $i$, $j$, $k$, in the same way as analyses of anisotropic flow~\cite{Adler:2003kt}, in order to suppress correlations due to decay kinematics and other ``nonflow'' effects. 
The skewness is likely to be less affected by nonflow effects than the variance as it is a higher-order cumulant~\cite{Borghini:2000sa}, but rapidity gaps can be easily implemented~\cite{DiFrancesco:2016srj}.

\section{Simple model of density fluctuations}
\label{s:sources}

In this appendix, we present an explicit application of the perturbative approach of Sec.~\ref{s:perturbative} by working out a simple example, and we assess its validity by showing the comparison between perturbative results and exact results coming from a Monte Carlo calculation. 

\subsection{Identical sources}
\label{s:id}
We model the entropy density at the beginning of the hydrodynamic evolution as the sum of $N$ identical contributions~\cite{Bhalerao:2011bp}, in the spirit of the Glauber modeling~\cite{Miller:2007ri}:
\begin{equation}
\label{sourcemodel}
  s(x)=\sum_{i=1}^N \Delta(x-r_i),
  \end{equation}
where $r_i$ are the positions of  ``sources'', whose positions in the transverse plane are independent random variables with a probability distribution $p(r_i)$, and $\Delta(x)$ is a narrow peak centered around the origin. 
The total entropy is $\int_x s(x)=N\int_x\Delta(x)$. 
Therefore, fixing the total entropy amounts to fixing the number of sources, $N$. 

The $n$-point functions of this model can be evaluated explicitly in terms of $N$, $p(x)$ and $\Delta(x)$~\cite{Gelis:2019vzt}. 
The 1-point function is:
\begin{equation}
\label{1ptsource}
\langle s(x)\rangle=N\int_r p(r)\Delta(x-r),
\end{equation}
while the 2-point function is:
\begin{align}
\label{2ptsource}
\nonumber \langle \delta s(x_1)\delta s(x_2)\rangle&=N \int_r p(r)\Delta(x_1-r)\Delta(x_2-r) \\ 
&-N \int_r p(r)\Delta(x_1-r) \int_{r'} p(r')\Delta(x_2-r').
\end{align}

If the width of the function $\Delta(r)$ is much smaller than the scale over which $p(x)$ varies, one can neglect the variation of $p(x)$ across the extension of the source, and these equations simplify to:
\begin{equation}
\label{1ptsource2}
\langle s(x)\rangle=Np(x)\int_r \Delta(r),
\end{equation}
and 
\begin{eqnarray}
\label{2ptsource2}
\langle \delta s(x_1)\delta s(x_2)\rangle&&=N p(x_1)\int_r \Delta(x_1-r)\Delta(x_2-r)\cr &&-N
 p(x_1)p(x_2)\left(\int_r \Delta(r)\right)^2.
\end{eqnarray}
Note that the latter equation is a specific case of Eq.~(\ref{2pointc}), with
\begin{equation}
\label{kappa2}
  \kappa_2(x)=N p(x)\left(\int_r \Delta(r)\right)^2,
\end{equation}
which amounts to assimilating the sources to Dirac delta peaks.
A similar calculation~\cite{Gronqvist:2016hym} shows that the 3-point function has the same form as in Eq.~(\ref{3pointc}), with
\begin{equation}
\label{kappa3}
  \kappa_3(x)=N p(x)\left(\int_r \Delta(r)\right)^3.
\end{equation}
Note that both $\kappa_2(x)$ and $\kappa_3(x)$ depend on the integral of $\Delta(x)$ over the plane and, thus, are independent of the actual shape of this function. This confirms somewhat more explicitly the previous argument that the variance and the skewness are indeed not sensitive to short-scale structures.

\subsection{Gaussian density profile, constant $c_s$} 
\label{s:cs}

To move forward, we need to specify the functional form of $p(r_i)$ and an equation of state.
To obtain compact analytic expressions, we consider for simplicity 
that the distribution of sources in the transverse plane is Gaussian: 
\begin{equation}
\label{gaussianprofile}
  p(x)=\frac{1}{\pi\sigma^2}\exp\left(-\frac{x^2}{\sigma^2}\right).
\end{equation}
Then, according to Eq.~(\ref{1ptsource2}), the average entropy density profile is also Gaussian form. The energy density requires the knowledge of the equation of state. For simplicity, we consider a power-law equation of state:
\begin{align}
\label{blackbody}
\nonumber T&=s^{c_s^2},  \\
\epsilon&=\frac{s^{1+c_s^2}}{1+c_s^2},
\end{align}
where $c_s^2$ is the velocity of sound. 
At early times (or high temperatures), $c_s^2\simeq \frac{1}{3}$ in hydrodynamic calculations using the lattice QCD equation of state. 

We can thus proceed to the evaluation of the average temperature $\bar T$.
If we denote by $T_0$ the temperature in the center, the average entropy density and the corresponding temperature profiles are given by: 
\begin{align}
  \label{avsind}
\nonumber \langle s(x)\rangle&=T_0^{1/c_s^2}\exp\left(-\frac{x^2}{\sigma^2}\right) , \\
T(x)&=T_0\exp\left(-\frac{c_s^2 x^2}{\sigma^2}\right).
\end{align}
Identifying the first of these equations with Eq.~(\ref{1ptsource2}), one obtains:
\begin{equation}
  \int_r\Delta(r)=\frac{\pi\sigma^2 T_0^{1/c_s^2}}{N}.
\end{equation}
This expression can be used to express $\kappa_2(x)$ and $\kappa_3(x)$, defined by Eqs.~(\ref{kappa2}) and (\ref{kappa3}), as  a function of $N$, $c_s$ and $T_0$. 
Equation~(\ref{defbarT}) then gives: 
\begin{equation}
\label{barTgauss}
  \bar T=\frac{T_0}{1+c_s^2}.
\end{equation}

Finally, we can evaluate the mean, the variance, and the skewness of the initial energy, $E_0$, analytically using Eqs.~(\ref{E1}), (\ref{varE3}), (\ref{deltae33}), and (\ref{deltae34}). One obtains:
\begin{align}
\nonumber \langle E_0\rangle &=\frac{\pi \sigma^2}{(1+c_s^2)^2} T_0^{1+c_s^{-2}}, \\
\nonumber  \langle \delta E_0^2\rangle&= \frac{1}{N}\frac{\left(c_s^2(1+c_s^2)\right)^2}{(1+2c_s^2)} \langle E_0\rangle ^2, \\
\nonumber \langle \delta E_0^3\rangle_3&= - \frac{1}{N^2} (2-2c_s^2)\frac{\left(c_s^2(1+c_s^2)\right)^3}{(1+5c_s^2+6c_s^4)} \langle E_0\rangle^3, \\
\langle \delta E_0^3\rangle_4&= \frac{1}{N^2} (3+3c_s^2+6c_s^4)\frac{\left(c_s^2(1+c_s^2)\right)^3}{(1+5c_s^2+6c_s^4)}  \langle E_0\rangle^3.
\end{align}
The variance and the skewness are proportional to $1/N$ and $1/N^2$, respectively, as anticipated from the discussion in Sec.~\ref{s:dimensionless}. 
The two contributions to the skewness in Eq.~(\ref{deltae3dec}) are of the same order of magnitude. The first is negative while the second is positive, and larger in magnitude for any value of $c_s^2$. Note that, for a typical speed of sound, $c_s^2=1/3$, we find that $\langle \delta E_0^3\rangle_4$ is larger than $\langle \delta E_0^3\rangle_3$ by a factor 3. The positive term thus \textit{dominates}. This is a clear indication that the positive skewness of $E_0$ fluctuations is generic, and that one can safely expect to observe it in any model of the initial state.

Finally, the relative standard deviation, the standardized skewness, and the intensive skewness are given, respectively, by:
\begin{align}
\label{pertresults}
\nonumber \frac{\sqrt{ \langle \delta E_0^2\rangle} }{\langle E_0\rangle} &=\frac{1}{\sqrt{N}} \frac{c_s^2(1+c_s^2)}{\sqrt{1+2c_s^2}}, \\
\nonumber \gamma_{E_0}&= \frac{1}{\sqrt{N}} \left(1+2c_s^2\right)^{3/2}, \\
\Gamma_{E_0}&=\frac{(1+2c_s^2)^2}{c_s^2(1+c_s^2)}.
\end{align}
A few comments are in order. As anticipated in the discussion at the end of Sec.~\ref{s:var}, the relative fluctuation of $E_0$ is roughly proportional to $c_s^2$. 
The intensive skewness is independent of $N$, and inversely proportional to $c_s^2$. 
For $c_s^2=\frac{1}{3}$,  its value is $6.25$, which is actually close to the intensive skewness of the more sophisticated \trento{} calculation presented in Fig.~\ref{fig:hydroskew}(b).

\begin{figure}[t]
    \centering
    \includegraphics[width=.8\linewidth]{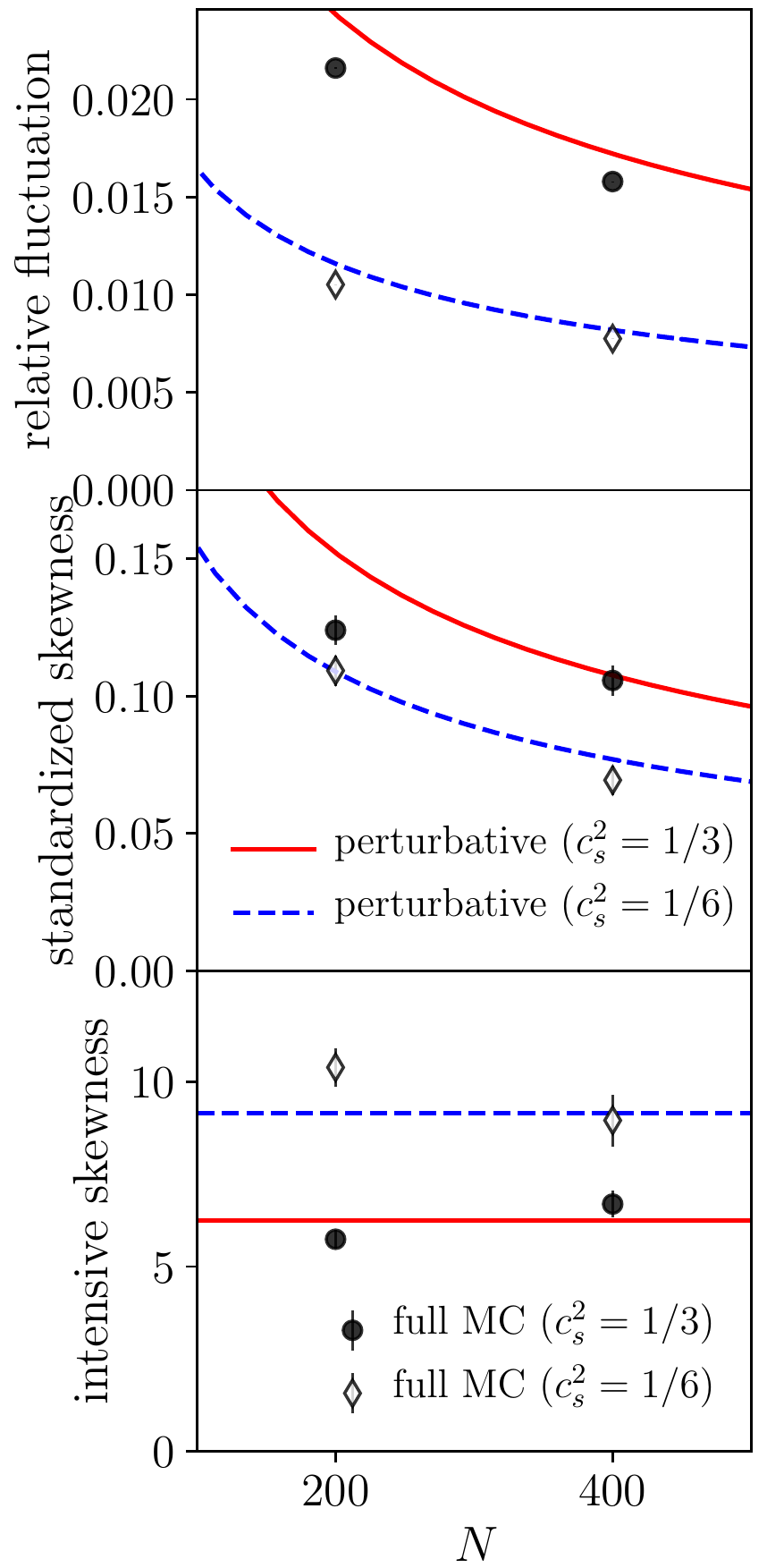}
    \caption{(Color online) Symbols: Results of Monte Carlo (MC) simulations (see text). Lines: leading order perturbative expression given by Eq,~(\ref{pertresults}). 
    Panels (a), (b) and (c) display the quantities corresponding to the three lines of these equations, respectively: relative standard deviation, standardized skewness, intensive skewness.
    The speed of sound is  $c_s=1/\sqrt{3}$ for the closed symbols and solid lines, while $c_s=1/\sqrt{6}$ for the open symbols and dashed lines. }
    \label{fig:montecarlo}
\end{figure}

\subsection{Monte Carlo calculations}
We check now the validity of the perturbative results by carrying out Monte Carlo simulations. To reproduce the model outlined in the previous section, the only additional ingredient to specify is the shape of a single source, $\Delta(x)$, appearing in Eq.~(\ref{sourcemodel}). One can use any function whose integral over the transverse plane is finite, since the final results do not depend on this choice, as argued previously. 
For simplicity, we choose a Gaussian:
\begin{equation}
\label{gaussianpeak}
 \Delta(x)\propto\exp\left(-\frac{x^2}{w^2}\right).
\end{equation}
The validity of the perturbative calculation relies on two conditions. 
First, the width of $\Delta(x)$, $w$, must be small compared to the typical transverse extent of one event (as determined by the positions of $N$ sources), which is in turn proportional to $\sigma$ in Eq.~(\ref{gaussianprofile}).
Second, the standard deviation of the entropy density at a given point, obtained as the square root of Eq.~(\ref{2ptsource2}) after setting $x_2=x_1$, must be smaller than the average density (\ref{1ptsource2}) at the same point, in order for the Taylor expansion in Eq.~(\ref{order2}) to be valid. Since in this source model the fluctuation of local quantities are determined by the density of sources at a given point, this is naturally a condition on the value of $N$.
In formulas, the conditions we need to fulfill are: 
\begin{align}
\label{conditions}
\nonumber \frac{w}{\sigma}&\ll 1, \\
\frac{\sigma}{w\sqrt{N}}&\ll 1.
\end{align}
We simply define $w$ by:
\begin{equation}
w=N^{-1/4}\sigma,
\end{equation}
so that both conditions (\ref{conditions}) are satisfied in the limit $N\gg 1$.

We generate a large number of Monte Carlo events. 
For each event, we sample the positions of $N$ sources, where $N$ is the same for all events, according to the distribution (\ref{gaussianprofile}). 
The initial entropy density in the event is then defined by Eqs.~(\ref{sourcemodel}) and (\ref{gaussianpeak}). 
We then compute the corresponding energy density, $\epsilon(x)$, using the equation of state, Eq.~(\ref{blackbody}). 
Since the equation of state is scale invariant, the final results are independent of the normalization constant in Eq.~(\ref{gaussianpeak}).  
We carry out two sets of calculations, using two different values of the speed of sound: $c_s^2=\frac{1}{3}$ corresponding to the quark-gluon plasma at high temperature, and a value twice smaller, in order to check that the analytic results capture the dependence of fluctuation observables on $c_s$. 
The total energy, $E_0$, is evaluated by integrating the energy density, $E_0\equiv\int_x\epsilon(x)$. 
Its cumulants (mean, variance, skewness) are finally evaluated by averaging over the ensemble of events. 

Figure~\ref{fig:montecarlo} displays our results for the relative fluctuation, the standardized skewness, and the intensive skewness, together with the perturbative results of Eqs.~(\ref{pertresults}). 
Agreement is not perfect, which shows that a leading-order perturbative calculation is not accurate enough even with a few hundred sources. 
Nevertheless, the perturbative results capture the order of magnitude and the dependence on $c_s^2$: 
In particular, Monte Carlo results confirm that a softer equation of state results in narrower fluctuations, with a larger intensive skewness.

\end{document}